\documentclass[11pt]{article}
\usepackage[utf8]{inputenc}
\usepackage{graphicx}   
\usepackage{amsmath, amssymb}  
\usepackage{hyperref}   
\usepackage{geometry}   
\usepackage{textgreek}  
\usepackage{natbib}     
\usepackage{doi}        
\usepackage{float}      

\title{Hallucinations in AlphaFold 3 for Intrinsically Disordered Proteins with disorder in Biological Process Residues}
\author{Shreya Gopalan; Sundaraparipurnan Narayanan \\[0.5em]
\small AI Tech Ethics \\
\small \texttt{shreya.gopalan@aitechethics.com}
}
\date{} 

\usepackage[utf8]{inputenc}
\usepackage[T1]{fontenc}
\usepackage{hyperref}
\usepackage{textgreek}
\usepackage{url}
\usepackage{booktabs}
\usepackage{amsfonts}
\usepackage{nicefrac}
\usepackage{microtype}
\usepackage{xcolor}

\begin{document}

\maketitle

\begin{abstract}Protein structure prediction has advanced significantly with the introduction of AlphaFold3, a diffusion-based model capable of predicting complex biomolecular interactions across proteins, nucleic acids, small molecules, and ions. While AlphaFold3 demonstrates high accuracy in folded proteins, its performance on intrinsically disordered proteins (IDPs)—which comprise 30–40 percent of the human proteome and play critical roles in transcription, signaling, and disease—remains less explored. This study evaluated AlphaFold3’s predictions of IDPs with a focus on intrinsically disordered regions (IDRs) using 72 proteins curated from the DisProt database. Predictions were generated across multiple random seeds and ensemble outputs, and residue-level pLDDT scores were compared with experimental disorder annotations. Our analysis reveals that 32 percent of residues are misaligned with DisProt, with 22 percent representing hallucinations where AlphaFold3 incorrectly predicts order in disordered regions or vice versa. Additionally, 10 percent of residues exhibited context-driven misalignment, suggesting that AlphaFold3 implicitly incorporates stable structural assumptions. Importantly, 18 percent of residues associated with biological processes showed hallucinations, raising concerns about downstream implications in drug discovery and disease research. These findings highlight the limitations of AlphaFold3 in modeling IDRs, the need for refined hallucination metrics beyond the pLDDT, and the importance of integrating experimental disorder data to improve the prediction reliability. 
\end{abstract}

\section{Introduction}

\subsection{Significance of Protein Structure Prediction and AlphaFold 3 advancements}

Protein structure prediction is of great importance because of its varied downstream applications in drug discovery and disease studies, along with the economic challenges of using experimental techniques to determine protein structure. AlphaFold3, developed by Google DeepMind in collaboration with Isomorphic Labs, is a transformative advancement in biomolecular (protein) prediction that allows progress in drug design and therapeutics \cite{desai2024review}. Advancements include a diffusion-based architecture capable of predicting complex biomolecular interactions, including proteins, nucleic acids, small molecules, ions, and modified residues \cite{abramson2024accurate}. AlphaFold is consistently evolving with widespread adoption in the domains of structural biology, and the open-sourcing of AlphaFold3 in November 2024 is a significant contribution to research progress. In addition, its predictions have been consistently validated against experimental studies \cite{kovalevskiy2024alphafold}.

\subsection{Intrinsically Disordered Proteins (IDPs) and Regions (IDRs)}

The success of AlphaFold 3 and its earlier versions in folded proteins is discussed while expressing its accuracy; however, the accuracy of intrinsically disordered proteins is limited. Intrinsically disordered proteins (IDP) are proteins which lack a stable three-dimensional structure at physiological conditions but are fully functional. The flexibility in the intrinsically disordered regions (IDRs) of the IDP offers the scope to interact with multiple targets \cite{sato2022biological}. One of the fundamental aspects of understanding is that disorders are not necessarily  binary in nature, and may be represented based on context-dependent diverse properties and functions. These properties, when observed in a specific context, exhibit variability in parameters such as pH, localization, binding, and post-translational modifications \cite{aspromonte2024disprot}. 

The biological relevance of IDPs and Intrinsically Disordered Regions (IDR) is being studied further, considering their prevalence in the eukaryotic genome (60\%). Furthermore, IDP’s play an indispensable role in biological processes such as transcription, translation, cell cycle, and signaling, and in turn, diseases. Intrinsically disordered proteins (IDPs), comprising 30–40\% of the proteome, are critical in diseases including neuro-degenerative disorders and cancer. For instance, 80\% of human cancer-associated proteins have long IDRs (e.g., p53 contains 50\% IDR in its sequence) \cite{wallin2017idps}\cite{lermyte2020idps}\cite{sato2022biological}. \cite{coskuner2024alphasyn} exhibited that neither AlphaFold2 nor AlphaFold3 did not fully capture structural characteristics of α-synuclein, an intrinsically disordered protein. Earlier studies have identified that Alphafold 2 demonstrates promising results in IDR, specifically with reference to conditionally folding binding regions/residues \cite{piovesan2022idrs}. Researchers have expressed some possible limitations in terms of chirality determination, conformation, modelling experimental parameters, and predicting structural flexibility through AF3 \cite{krokidis2025overview}\cite{fang2025opportunity}\cite{abramson2024accurate}. However, there is no extensive research that specifically analyses AlphaFold3’s performance against that of IDR. AlphaFold3 has introduced a diffusion-based probabilistic model, which, while improving the prediction efficiency of many biomolecules and their interactions, has also increased the rate of hallucinations \cite{abramson2024accurate}\cite{fang2025opportunity}.

\subsection{Research Objectives and Scope}

This research attempts to conduct such an analysis of IDP’s with a focus on IDR’s that have an impact on biological processes, with a focus on hallucinations that arise in predicting disordered regions. Further, given the learnings from AlphaFold2 research studies and AlphaFold3 diffusion-based ensemble architecture, this research attempts to cover implications of reproducibility and accuracy with reference to context-dependent IDRs.

\section{Methodology}

\subsection{DisProt Database}

DisProt is a database that links the structural and functional information of intrinsically disordered proteins \cite{aspromonte2024disprot}. DisProt is the primary database manually curated and annotated with experimental results for IDPs and  IDRs and has evolved with over 3000 IDPs. DisProt contains annotations relating to functional and structural aspects of disorders like disorder state, structural transition, biological processes, and molecular function based on experimental validation

\subsection{Protein Selection Criteria}

This research leverages the DisProt database to identify relevant IDPs for the research and associated information. The study selected 72 proteins for this study from the DisProt database by identifying proteins with Bio-Process and having IDRs greater than 75\% overall.

\subsection{AlphaFold3 Prediction Setup}

The AlphaFold server generated predictions for the identified IDPs. In order to understand the five ensemble variabilities of AlphaFold3, we experimented with seed variations to observe the extent of variability in predictions for IDRs. To that end, three types of seeds were provided as input while generating the predictions on the AlphaFold Server as follows: (1) No seed was provided manually. The AlphaFold server automatically attributes a random seed. (2) 5 as seeds and (3) 1234567890 as seeds. This approach resulted in three seed variations and five ensemble outputs for each, thereby having 15 model prediction outputs to evaluate reproducibility across prediction runs for each identified IDP.

\subsection{Analysis of pLDDT Scores and Variability}

The CIF files from the predictions were used to extract the pLDDT scores from the B-factor field programmatically. The pLDDT is a confidence score that indicates the reliability of the structural predictions for each residue, with scores above 70 typically considered to indicate high confidence \cite{williams2025lowplddt}. Furthermore, pLDDT scores represent protein disorder with a strong correlation with IDRs \cite{bruley2022digging}. 

The pLDDT scores were parsed to obtain the residue-level scores for each protein. These scores were compared for each seed to determine the seed-based variation for each protein.

pLDDT scores greater than 70 were considered as ‘ordered’ and less than 70 were considered as ‘disordered’ in correlation with the DisProt database.
An analysis of the variance in predictions based on seed-based input variations for the identified IDP in determining the variability influenced by different seed-based protein predictions. 

\subsection{Classification of Predictions}

A systemic comparison between the IDRs (at a residue level) for each protein from the DisProt database and the predictions (ordered/disordered using pLDDT scores) supported the determination of how well the AlphaFold3 protein predictions align with the ordered/disordered references from DisProt. 

Based on the above analysis, proteins were classified into three types: aligned (when AF3 and DisProt are aligned), hallucination (when AF3 shows order but DisProt shows disorder for regions that are not involved in structural transition, and when AF3 shows disorder while DisProt shows order), and possible context-driven misalignment (when AF3 shows order for experimentally proven disordered residues with structural transition). 45 proteins were not considered for context based due to the absence of the structural transition annotation for these proteins.

\subsection{Context-Based Interaction Modeling}

Of the 27 proteins with context-driven misalignment, 17 proteins were modelled with other biomolecules with which they typically interact to determine the basis for these hallucinations. 

The other 10 proteins could not be modelled because the interacting components could not be modelled using AlphaFold3 (e.g., environment-induced disorder).

For the 17 proteins, potential interacting biomolecules were identified (based on structural transition literature available in DisProt) and predictions were gathered (adopting the same 3 seed approach) from AlphaFold server, the associated pLDDT were extracted and were classified as ‘ordered’ or ‘disordered’ based on a threshold pLDDT score of 70 (as referred above). A comparative analysis was performed for the prediction results similar to the above process.

\section{Results}
\subsection{Variance based on seed input}
The analysis revealed a lack of significant variance in AlphaFold3 predictions when varying seed inputs and using ensemble models, with consistent pLDDT scores across seeds (e.g., no seed, 5, 1234567890). This suggests the ensemble approach may not effectively capture structural variability in IDRs across multiple runs.

\subsection{Alignment with DisProt Annotations}

Initial analysis of alignment of AlphaFold3 predictions as ordered/disordered was done by comparing the ordered/disordered as documented in DisProt for the identified 72 proteins. More than 50\% of the proteins had less than 70\% alignment with DisProt. Furthermore, the results showed that 68\% of the residues aligned with DisProt and 32\% did not. 

In addition, 7.6\% of the aligned residues were involved in structural transition, which implies that AF3 predicted the native structural state of the residues without assuming an unrepresented context, termed as incidental alignment. For instance, Apolipoprotein AI (DP04271) exhibited order in only one residue, against 242 residues (91\%) that were recorded as disordered in DisProt.

\begin{table}[H]
    \centering
    \includegraphics[width=0.95\linewidth]{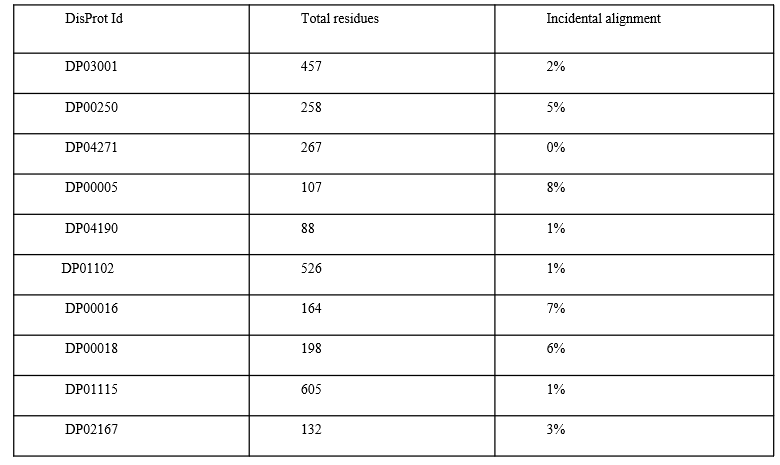}
    \caption{Percentage of incidental alignment in selected proteins}
    \label{tab:1}
\end{table}

\subsection{Hallucination Analysis}
This study identified two types of hallucinations. (a) DisProt shows order in the residue, but AF3 predicts the residue with low confidence, implying disorder. (b) DisProt shows disorder, and AF3 predicts it with high confidence without a structural transition potential.  

\cite{abramson2024accurate} noted that hallucinations (order) in disordered regions were typically flagged with low confidence scores. However, our study found instances where high confidence scores were assigned to disordered regions and ordered structures were predicted with low confidence, both of which were identified as hallucinations in this study.

22\% of the residues were hallucinations ((a)+(b)). For instance, proteins such as the nuclear export protein (DP00871), calreticulin (DP00333), functional amyloid subunits FapB and FapC (DP04435 and DP04433, respectively), and von Hippel-Lindau disease tumor suppressor (DP00287) exhibited hallucinations with more than 70\% of residues.

\begin{figure}[H]
    \centering
    \includegraphics[width=0.95\linewidth]{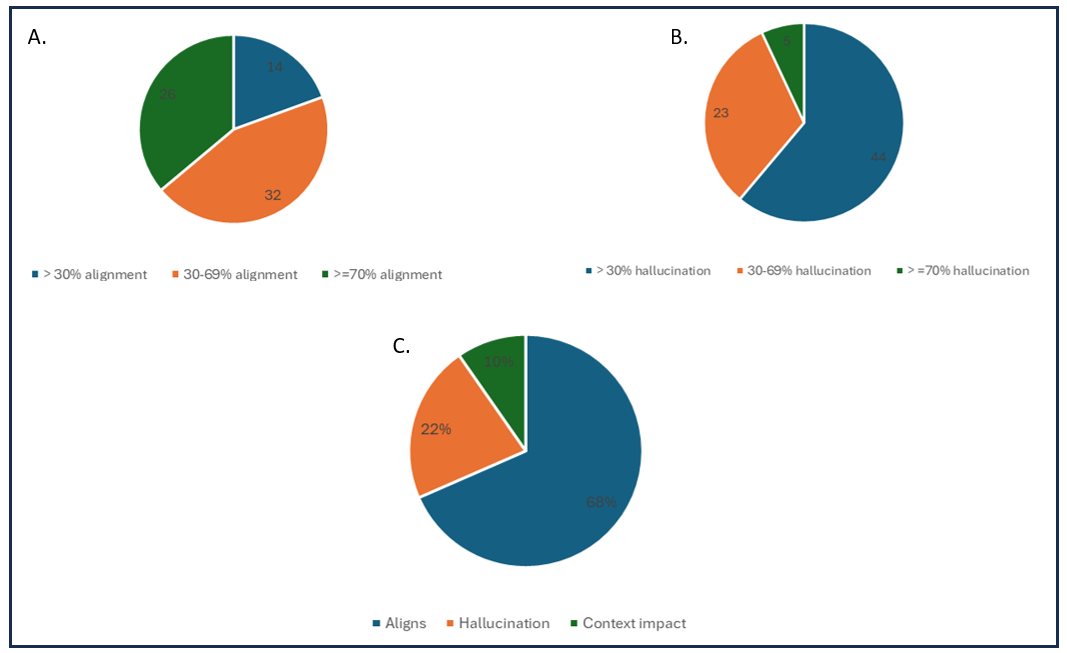}
    \caption{A. Number of proteins with percentage of alignment B. Number of proteins with percentage of hallucinations C. Overall alignment of residues with DisProt}
    \label{fig:1}
\end{figure}

\subsection{Context-Driven Misalignment}
Instances where AF3 predicted a residue with high confidence when the DisProt data showed disorder and is known to have structural transition were identified for subsequent validations. Context-based interactions were identified, and predictions were run for these proteins on AF3, considering the potential context-assumed alignment against stable experimental evidence. For instance, Disprot id DP04016 exhibited ordered in 82\%, while it is an IDP with 100\% of residues to be disordered, indicating potential assumption of context. AlphaFold 3 has mentioned that there are potential instances wherein the predictions showed inclinations to represent closed state confirmation even when the native confirmation is in an open state \cite{abramson2024accurate}.

A total of 27 proteins (10\% of the overall residue counts considered for review) were found to have residues with context-assumed hallucinations, of which 17 proteins were modelled with other biomolecules with which they typically interact to determine the basis for these misalignments.
For instance, proteins such as seed maturation protein (DP01442) and uncharacterized protein (DP03738) fold due to environment-induced stress conditions, and proteins such as LEA proteins (DP04016, DP01858, DP04018, DP04019) and ICP47 protein (DP04190) interact with small molecules such as ethylene glycol and sodium dodecyl sulfate, which cannot be represented in AF3. Other proteins like temporin - 1Tl (DP03818), interact with lipids that cannot be modelled in AF3. The Tat protein (DP01087) was modelled with an Fab molecule generated from an antigen-induced mouse, for which the antibody sequence could not be retrieved.

Of the remaining 17 proteins, a few proteins paratox (DP04243) and temporin - 1Tl (DP03818) showed 100\% possibility of context-assumed hallucinations, followed by proteins such as antitoxin phd (DP00288), seed maturation protein (DP01442), and Late Embryogenesis Abundant (LEA) family protein (DP04019), indicating possible order due to inherently context-assumed predictions (>80\%). \cite{abramson2024accurate} suggests that the AF3 process takes into account possible stable confirmation at residue level and attempts to align the folds to form such stabler confirmations. 

Comparison of predictions (ordered/disordered) in the native state with predictions in the context-based state from AF3 revealed that the latter results aligned with the former in 89\% of residues (15 out of 17 proteins with more than 80\% alignment with the previous run). The only exceptions were alpha synuclein (DP00070) with a match of 38\% and tat protein (DP00929) with an alignment of 73\% of residues. This raises questions on whether context-based consideration has any significant impact on proteins. Alternatively, it also brings into question whether AF3 predictions are inherently influenced by context assumptions. 

Comparison of predictions (ordered/disordered) based on DisProt with predictions with context-based state from AF3 revealed that the later results aligned with the former in 68\% of residues. 
Six out of 17 proteins considered for this analysis had over 80\% of residues aligned with DisProt, and in 11 out of 17 cases, 65\% of residues aligned with DisProt, questioning if the prior predictions were influenced by context-based considerations that AF3 takes into account. 

The exceptions to this trend were the following proteins: antitermination protein N (DP00005), minor curli subunit (DP03852), major curli subunit (DP03853), cholera enterotoxin subunit A (DP00250), and paratox (DP04243). While this review and analysis represents the study of whole protein-related predictions, a specific study focused on IDRs may provide more insights into specific proteins across their context-based interactions. 
However, the research also noted a protein alpha synuclein (DP00070) that exhibited increased alignment with DisProt in the Context-based prediction run, indicating that the context-based interaction made the protein prediction even more disordered than the prior native AF3 run. This contradicts the role of interacting proteins in transitioning disordered residues to order \cite{cartelli2016synuclein}.
Another instance of protein interaction is that of the cholera enterotoxin subunit A and its interaction with protein disulfide isomerase (PDI), which functions as a dual agent to disassemble and reassemble the enterotoxin subunits based on the redox environment \cite{tsai2001pdi}. Such instances cannot be modelled in the AlphaFold server currently. Our analysis showed an alignment of 84\% with the native state AF3 run and 21\% with DisProt, which might suggest an inclination toward the oxidation state of PDI, which leads to folding of the enterotoxin by AF3.

\subsection{Impact on Biological Processes}
This study utilized the biological process annotation available in the DisProt database to determine the impact of hallucinations in the residues experimentally validated to contribute to biological processes.
Overall, 18\% of the residues involved in the biological processes showed hallucinations. 

Of the 72 proteins, 30\% did not show hallucinations at segments known to have a biological context. Eighteen \% of the proteins showed hallucinations, ranging from 1 to 10\%. 8\% of the proteins show a higher degree of hallucination (60-80\%) at the regions involved in biological processes. For instance, proteins that are functional amyloid subunits, FapB and FapC (DP04435 and DP04433), show 75\% and 77\% hallucinations, respectively. The von Hippel-Lindau disease tumor suppressor shows 73\% hallucinations in residues attributed to biological processes. Protein Vpr (DP03543), an active regulator of SIRT1 (DP03988), and GRASP65 homolog protein 1 (DP02544) were found to have hallucinations in the range of 60\% to 65\% at biologically relevant residues.

\section{Discussion}

\subsection{Implications of Hallucinations and Misalignments}
The human proteome has approx 20000 protein-coding genes (based on current UniProt/Ensembl data) in its canonical form. Of which 30-40\% of the proteomes (approximately 8000) are intrinsically disordered. The number of IDP’s documented in DisProt is approximately 3000; hence, the above analysis needs to be considered from the lens of both known IDPs with research and annotations on experimental validations and other IDPs without experimental confirmations (remaining approximately 5000). 

Hallucinations and misalignments in known IDP’s are comparable and validatable in downstream uses including drug discovery, disease diagnostics, and molecule research. However, such opportunities to identify or analyze IDPs without experimental confirmation may not clearly exist. Hence, the learnings and observations from known IDPs provide referrable guidance for downstream considerations of IDPs without experimental confirmation. 
From this lens, hallucination and misalignment, in general, could lead to misleading diagnostics or drug targets, delaying therapies and inflating costs due to the need for more experiments. 

Specifically, hallucinations in known IDPs represent the need for specific metrics to measure hallucinations or identify potential hallucinations. While pLDDT is considered representative for this purpose, a high pLDDT in disordered regions \cite{abramson2024accurate} expresses overconfidence of the prediction model, exploring alternative metrics to measure hallucination effectively in such predictions, as we evolve the process of documenting experimental validation of IDPs. In addition, such hallucinations and misalignment may misguide therapeutic targets and biomarker discovery approaches as part of drug discovery or disease-target explorations and diagnostics. In addition, hallucinations and misalignments, especially in IDPs where experimental confirmations are fewer or need more research, leave severe implications in downstream adoption for target identification for drugs.

\subsection{Limitations of Current Analysis}
The generalizability of the study is constrained by its limited sample size (72 out of 3000+ DisProt proteins) and exclusive focus on DisProt-annotated proteins, omitting uncharacterized intrinsically disordered proteins (IDPs). Additionally, the reliance on seed-based variability analysis may not fully capture AlphaFold3's ensemble diversity, and AlphaFold3's inability to model certain interactions (e.g., lipids and environment-induced effects) poses a significant limitation. The simplification of disorder-order classification through pLDDT thresholds (>70)  requires revision, as generic thresholds for determining ordered or disordered could be misleading contextually.

\section{Conclusion}
This study quantitatively assessed AlphaFold3's performance on intrinsically disordered proteins (IDPs), revealing significant variability in predictions, particularly concerning non-deterministic hallucinations and context-driven misalignments. These findings indicate that while AlphaFold3 excels in folded protein prediction, its application to IDPs necessitates refined evaluation metrics beyond pLDDT, given instances of high confidence scores assigned to disordered regions. The observed practical implications of these discrepancies underscore the critical need for enhanced research and improvement of AF3 model feedback or continuous learning processes, especially as we extend these models to novel IDPs without existing experimental validation.

\subsection{Future Directions for IDP Prediction}
Future research should aim to expand the dataset to include a wider range of IDPs and intrinsically disordered regions (IDRs) for robust validation. Developing metrics beyond pLDDT is crucial for accurately detecting and quantifying hallucinations in IDP predictions. Integrating multi-omics data, such as pH and post-translational modifications, will enable context-aware predictions, and the enhancement of AlphaFold3 with iterative feedback from DisProt is an essential step towards reducing predictive hallucinations.

\subsection{Recommendations for Model Improvement}
Given that the ensemble approach did not provide significant variance across multiple seed runs for the identified IDPs, it may be necessary to understand how the ensemble approach could be used to identify hallucinations rather than collectively confident ensembles leading to hallucinations. In addition, a mechanism to allow for AF3 to perform such comparisons leveraging DisProt to enable the opportunity to improve model performance over period.

\bibliographystyle{plain}  
\bibliography{references}

@article{desai2024review,
  author    = {Desai, D. and Kantliwala, S. V. and Vybhavi, J. and Ravi, R. and Patel, H. and Patel, J.},
  title     = {Review of AlphaFold 3: Transformative Advances in Drug Design and Therapeutics},
  journal   = {Cureus},
  year      = {2024},
  month     = {jul},
  doi       = {10.7759/cureus.63646}
}

@article{abramson2024accurate,
  author    = {Abramson, J. and others},
  title     = {Accurate structure prediction of biomolecular interactions with AlphaFold 3},
  journal   = {Nature},
  volume    = {630},
  number    = {8016},
  pages     = {493--500},
  year      = {2024},
  month     = {jun},
  doi       = {10.1038/s41586-024-07487-w}
}

@article{kovalevskiy2024alphafold,
  author    = {Kovalevskiy, O. and Mateos-Garcia, J. and Tunyasuvunakool, K.},
  title     = {AlphaFold two years on: Validation and impact},
  journal   = {Proceedings of the National Academy of Sciences},
  volume    = {121},
  number    = {34},
  year      = {2024},
  month     = {aug},
  doi       = {10.1073/pnas.2315002121}
}

@article{sato2022biological,
  author    = {Sato, M.},
  title     = {Biological Significance of Intrinsically Disordered Protein Structure},
  journal   = {Chem-Bio Informatics Journal},
  volume    = {22},
  number    = {0},
  pages     = {26--37},
  year      = {2022},
  month     = {may},
  doi       = {10.1273/cbij.22.26}
}

@article{aspromonte2024disprot,
  author    = {Aspromonte, M. C. and others},
  title     = {DisProt in 2024: improving function annotation of intrinsically disordered proteins},
  journal   = {Nucleic Acids Res},
  volume    = {52},
  number    = {D1},
  pages     = {D434--D441},
  year      = {2024},
  month     = {jan},
  doi       = {10.1093/nar/gkad928}
}

@article{wallin2017idps,
  author    = {Wallin, S.},
  title     = {Intrinsically disordered proteins: structural and functional dynamics},
  journal   = {Res Rep Biol},
  volume    = {8},
  pages     = {7--16},
  year      = {2017},
  month     = {feb},
  doi       = {10.2147/RRB.S57282}
}

@article{lermyte2020idps,
  author    = {Lermyte, F.},
  title     = {Roles, Characteristics, and Analysis of Intrinsically Disordered Proteins: A Minireview},
  journal   = {Life},
  volume    = {10},
  number    = {12},
  pages     = {320},
  year      = {2020},
  month     = {nov},
  doi       = {10.3390/life10120320}
}

@article{coskuner2024alphasyn,
  author    = {Coskuner-Weber, O.},
  title     = {Structures prediction and replica exchange molecular dynamics simulations of $\alpha$-synuclein: A case study for intrinsically disordered proteins},
  journal   = {Int J Biol Macromol},
  volume    = {276},
  pages     = {133813},
  year      = {2024},
  month     = {sep},
  doi       = {10.1016/j.ijbiomac.2024.133813}
}

@article{piovesan2022idrs,
  author    = {Piovesan, D. and Monzon, A. M. and Tosatto, S. C. E.},
  title     = {Intrinsic protein disorder and conditional folding in AlphaFoldDB},
  journal   = {Protein Science},
  volume    = {31},
  number    = {11},
  year      = {2022},
  month     = {nov},
  doi       = {10.1002/pro.4466}
}

@article{krokidis2025overview,
  author    = {Krokidis, M. G. and others},
  title     = {AlphaFold3: An Overview of Applications and Performance Insights},
  journal   = {Int J Mol Sci},
  volume    = {26},
  number    = {8},
  pages     = {3671},
  year      = {2025},
  month     = {apr},
  doi       = {10.3390/ijms26083671}
}

@article{fang2025opportunity,
  author    = {Fang, Z. and others},
  title     = {AlphaFold 3: an unprecedent opportunity for fundamental research and drug development},
  journal   = {Precis Clin Med},
  volume    = {8},
  number    = {3},
  year      = {2025},
  month     = {jun},
  doi       = {10.1093/pcmedi/pbaf015}
}

@article{williams2025lowplddt,
  author    = {Williams, C. J. and Chen, V. B. and Richardson, D. C. and Richardson, J. S.},
  title     = {Categorizing prediction modes within low-pLDDT regions of AlphaFold2 structures},
  journal   = {bioRxiv},
  year      = {2025},
  month     = {jun},
  day       = {7},
  doi       = {10.1101/2025.06.06.658382}
}

@article{bruley2022digging,
  author    = {Bruley, A. and Mornon, J.-P. and Duprat, E. and Callebaut, I.},
  title     = {Digging into the 3D Structure Predictions of AlphaFold2 with Low Confidence: Disorder and Beyond},
  journal   = {Biomolecules},
  volume    = {12},
  number    = {10},
  pages     = {1467},
  year      = {2022},
  month     = {oct},
  doi       = {10.3390/biom12101467}
}

@article{cartelli2016synuclein,
  author    = {Cartelli, D. and Aliverti, A. and Barbiroli, A. and others},
  title     = {{$\alpha$}-Synuclein is a Novel Microtubule Dynamase},
  journal   = {Scientific Reports},
  volume    = {6},
  pages     = {33289},
  year      = {2016},
  doi       = {10.1038/srep33289},
  url       = {https://doi.org/10.1038/srep33289}
}

@article{tsai2001pdi,
  author    = {Tsai, B. and Rodighiero, C. and Lencer, W. I. and Rapoport, T. A.},
  title     = {Protein disulfide isomerase acts as a redox-dependent chaperone to unfold cholera toxin},
  journal   = {Cell},
  volume    = {104},
  number    = {6},
  pages     = {937--948},
  year      = {2001},
  doi       = {10.1016/S0092-8674(01)00289-6},
  pmid      = {11290330}
}
\small


\end{document}